\documentclass[lettersize,journal]{IEEEtran}
\usepackage{amsmath,amsfonts}
\usepackage{algorithmic}
\usepackage{algorithm}
\usepackage{array}
\usepackage[caption=false,font=normalsize,labelfont=sf,textfont=sf]{subfig}
\usepackage{textcomp}
\usepackage{stfloats}
\usepackage{url}
\usepackage{verbatim}
\usepackage{graphicx}
\usepackage{cite}
\usepackage{hyperref}

\usepackage{pifont}  % 为了使用 \ding{55} (叉号)
\usepackage{multirow}
\usepackage{xcolor}   % 用于颜色设置
\usepackage{colortbl}   % 表格行着色
\definecolor{headercolor}{RGB}{170,74,74} % 定义头部深红色
\definecolor{lightgray}{gray}{0.9}        % 浅灰色
\usepackage{booktabs}

\begin{document}

\title{Data Poisoning in Deep Learning: A Survey}

\author{Pinlong Zhao, Weiyao Zhu, Pengfei Jiao, Di Gao, Ou Wu
        % <-this % stops a space

\thanks{This work was partially supported by  the National Natural Science Foundation of China under Grant 62476191. (\textit{Corresponding author: Ou wu.})}% <-this % stops a space
\thanks{Pinlong Zhao, and Weiyao Zhu contributed equally.}
\thanks{Pinlong Zhao and Pengfei Jiao are with the School of Cyberspace, Hangzhou Dianzi University, Hangzhou 310018, China. E-mail: pinlongzhao@hdu.edu.cn, pjiao@hdu.edu.cn.}% <-this % stops a space
\thanks{Weiyao Zhu is with National Center for Applied Mathematics, Tianjin University, Tianjin, China, 300072. E-mail: wyzhu@tju.edu.cn.}
\thanks{ Di Gao and Ou Wu are with HIAS, University of Chinese Academy of Sciences, Hangzhou, China, 310024. E-mail: gaodi@ucas.ac.cn, wuou@ucas.ac.cn.}

}



\maketitle

\begin{abstract}
Deep learning has become a cornerstone of modern artificial intelligence, enabling transformative applications across a wide range of domains. As the core element of deep learning, the quality and security of training data critically influence model performance and reliability. However, during the training process, deep learning models face the significant threat of data poisoning, where attackers introduce maliciously manipulated training data to degrade model accuracy or lead to anomalous behavior. While existing surveys provide valuable insights into data poisoning, they generally adopt a broad perspective, encompassing both attacks and defenses, but lack a dedicated, in-depth analysis of poisoning attacks specifically in deep learning. In this survey, we bridge this gap by presenting a comprehensive and targeted review of data poisoning in deep learning. First, this survey categorizes data poisoning attacks across multiple perspectives, providing an in-depth analysis of their characteristics and underlying design princinples. Second, the discussion is extended to the emerging area of data poisoning in large language models(LLMs). Finally, we explore critical open challenges in the field and propose potential research directions to advance the field further. To support further exploration, an up-to-date repository of resources on data poisoning in deep learning is available at \url{https://github.com/Pinlong-Zhao/Data-Poisoning}.
\end{abstract}


\section{Introduction}
\IEEEPARstart{O}{ver} the past decade, machine learning, particularly deep learning, has made remarkable progress in the field of artificial intelligence (AI), driving transformative advancements across industries and society as a whole~\cite{lecun2015deep}. From image recognition~\cite{he2016deep,liu2022exploring} and speech processing~\cite{dong2018speech,kim2024prompt} to natural language understanding~\cite{vaswani2017attention,lewis2019bart,kenton2019bert}, deep learning models have achieved groundbreaking success in numerous applications, significantly enhancing the automation and precision of intelligent systems. Notably, the latest developments in large language models (LLMs) have demonstrated exceptional learning and reasoning capabilities~\cite{brown2020language,ramesh2022hierarchical,achiam2023gpt,chowdhery2023palm,abdin2024phi}, propelling AI towards higher levels of intelligence and even being considered a potential key to achieving Artificial General Intelligence (AGI). These advancements have been primarily driven by the availability of immense computational power and diverse, large-scale training datasets, which together form the foundation for the rapid development of modern artificial intelligence.

\begin{figure}[t] 
    \centering \includegraphics[width=0.8\linewidth]{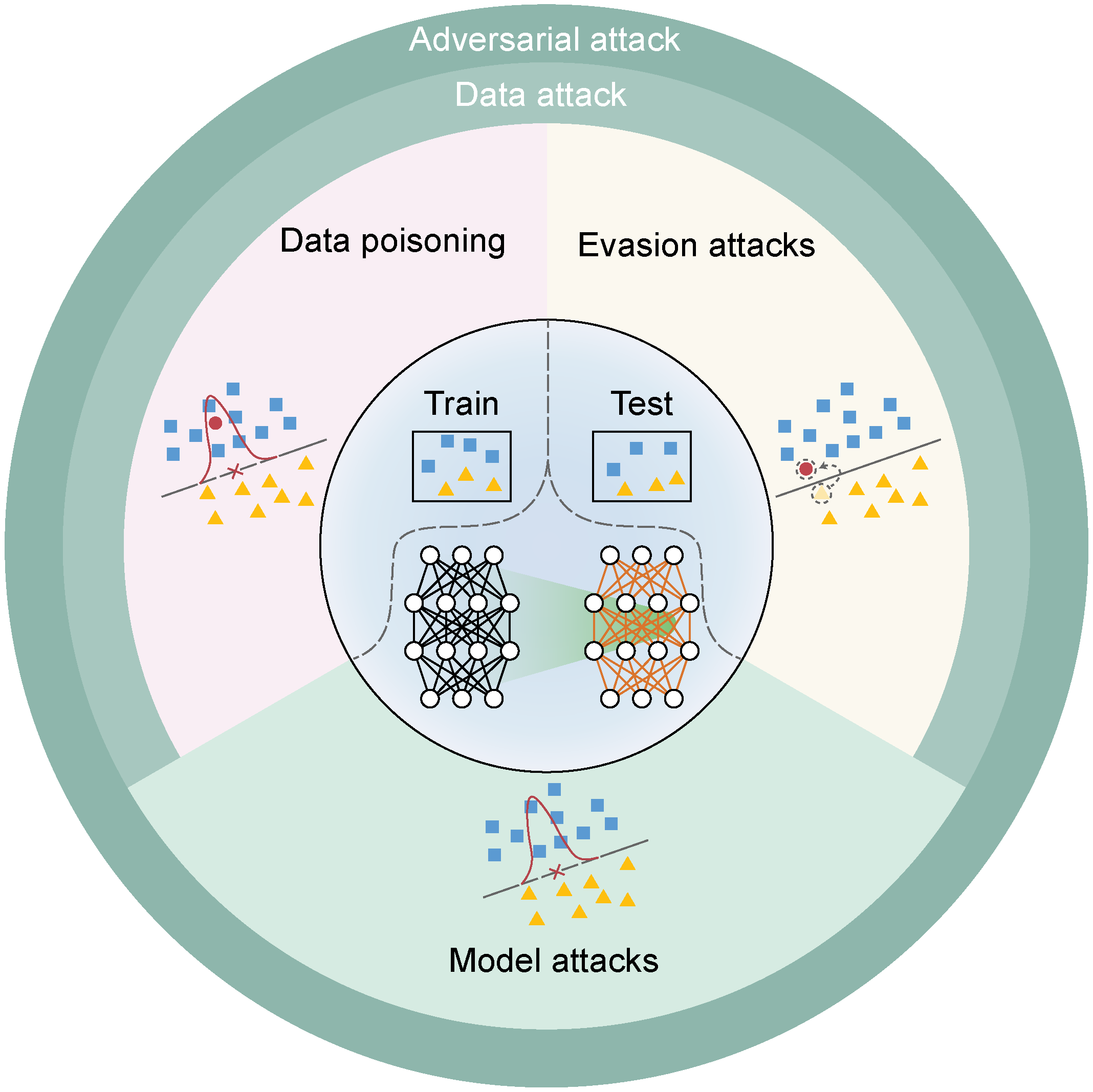}    \vspace{-0.12in}
    \caption{Types of Adversarial Attacks.}
    \label{fig1}
     \vspace{-0.1in}
\end{figure}

Despite these advancements, the increasing deployment of AI systems in critical domains such as healthcare, finance, and transportation has raised serious concerns about their security and reliability. The performance of AI models heavily relies on the quality of their training data and robustness to perturbations, yet this dependence also exposes them to a variety of potential security threats. By crafting sophisticated attack strategies, attackers can disrupt the normal functioning of AI systems, leading to erroneous decisions or even catastrophic failures in mission-critical tasks. Among these threats, adversarial attacks have been widely recognized as one of the most challenging risks facing AI today. Through manipulation of input data or exploiting model vulnerabilities, adversarial attacks can directly undermine model performance, posing severe threats to the safety and reliability of AI applications.

Adversarial attacks can be categorized into several types based on their methodologies, including model attacks~\cite{fang2020local,panda2022sparsefed}, evasion attacks (also known as adversarial examples)~\cite{akhtar2018threat,akhtar2021advances,chen2022adversarial}, and data poisoning attacks (also known as data poisoning or poisoning attacks)~\cite{ma2021poisoning,yerlikaya2022data,sun2022adversarial}, as illustrated in Fig.~\ref{fig1}. Model and evasion attacks typically target trained models. Evasion attacks primarily occur during the test phase, where adversarial examples are crafted to mislead predictions. Model attacks, on the other hand, aim to compromise the integrity of the model itself by tampering with its parameters, structure, or inserting hidden functionalities.  In contrast, data poisoning attacks, which occur during the training phase, are more covert and destructive, aiming to compromise the training model by injecting malicious samples into the training data. This manipulation can fundamentally degrade the model performance or alter its behavior. Thus, comprehensively understanding and mitigating data poisoning attacks is crucial for safeguarding the security and reliability of AI systems.

\begin{table*}[t!]
\centering
\caption{ The Comparison between Our Work and Existing Surveys}
\resizebox{\textwidth}{!}{
\small
\renewcommand{\arraystretch}{1.3} % 增加行距，使表格更美观
\begin{tabular}{c c c c c c c c c c c c c c c c c c c}
\hline
\multirow{2}{*}{\textcolor{headercolor}{Paper}} & 
\multicolumn{4}{c}{\textcolor{headercolor}{Scope}} & 
\multicolumn{7}{c}{\textcolor{headercolor}{Data Poisoning Attacks}} & 
\multicolumn{6}{c}{\textcolor{headercolor}{Data Poisoning Algorithms}} & 
\multirow{2}{*}{\textcolor{headercolor}{LLMs}} \\ 
\cmidrule(lr){2-5} \cmidrule(lr){6-12} \cmidrule(lr){13-18}
& \textcolor{headercolor}{TML} & \textcolor{headercolor}{DL} & \textcolor{headercolor}{A} & \textcolor{headercolor}{D} 
& \textcolor{headercolor}{AO} & \textcolor{headercolor}{AG} & \textcolor{headercolor}{AK} & \textcolor{headercolor}{ASt} & \textcolor{headercolor}{ASc} & \textcolor{headercolor}{AI} & \textcolor{headercolor}{AV}
& \textcolor{headercolor}{HA} & \textcolor{headercolor}{LP} & \textcolor{headercolor}{FSA} & \textcolor{headercolor}{BO} & \textcolor{headercolor}{IM} & \textcolor{headercolor}{GA}\\ 
\hline
\rowcolor{lightgray} Pitropakis 2019~\cite{pitropakis2019taxonomy} & \checkmark & \checkmark & \checkmark &   & 
  & \checkmark & \checkmark &   &   &   &   & 
 &   &   &   &   &   &   \\ 
Tahmasebian 2020~\cite{tahmasebian2020crowdsourcing} & \checkmark & \checkmark & \checkmark &   & 
  & \checkmark & \checkmark &   &   &   &   & 
\checkmark &   &   & \checkmark  &   &   &   \\ 
\rowcolor{lightgray} Ahmed 2021~\cite{ahmed2021threats} & \checkmark & \checkmark & \checkmark & \checkmark  & 
  &  &  &   &   &   &   & 
 &   &   &   &   &   &   \\ 
Ramirez 2022~\cite{ramirez2022poisoning} & \checkmark & \checkmark & \checkmark & \checkmark  & 
  &  & \checkmark &   &   &   &   & 
 &  \checkmark & \checkmark  &  \checkmark &   &  \checkmark &   \\ 
\rowcolor{lightgray} Fan 2022~\cite{fan2022survey} & \checkmark & \checkmark & \checkmark & \checkmark  & 
  & \checkmark &  &   &   &   &   & 
 &   &   &  \checkmark &   &   &   \\ 
Tian 2022~\cite{tian2022comprehensive} & \checkmark & \checkmark & \checkmark & \checkmark  & 
\checkmark  & \checkmark & \checkmark &   &   &   &   & 
 & \checkmark  &   &  \checkmark &   &   &   \\ 
\rowcolor{lightgray} Wang 2022~\cite{wang2022threats} & \checkmark & \checkmark & \checkmark & \checkmark  & 
\checkmark  & \checkmark & \checkmark &   &   &   &   & 
 &   &  \checkmark &  \checkmark & \checkmark  & \checkmark  &   \\ 
Goldblum 2022~\cite{goldblum2022dataset} & \checkmark & \checkmark & \checkmark & \checkmark  & 
  &  &  &   &   &   &   & 
 &  \checkmark &  \checkmark &  \checkmark & \checkmark  &   &   \\ 
\rowcolor{lightgray} Xia 2023~\cite{xia2023poisoning} & \checkmark & \checkmark & \checkmark & \checkmark  & 
  & \checkmark &  &   &   &   &   & 
 &   &   &   &   &   &   \\ 
Tayyab 2023~\cite{tayyab2023comprehensive} & \checkmark & \checkmark & \checkmark & \checkmark  & 
  &  & \checkmark &   &   &   &   & 
 &   &   &   &   &   &   \\ 
\rowcolor{lightgray} Cina 2023~\cite{cina2023wild} & \checkmark & \checkmark & \checkmark & \checkmark  & 
 \checkmark & \checkmark & \checkmark &   &   &   &   & 
 &  \checkmark &  \checkmark &  \checkmark &   &   &   \\ 
Cina 2024~\cite{cina2024machine} & \checkmark & \checkmark & \checkmark & \checkmark  & 
  & \checkmark &  &   &   &   &   & 
 &   &   &   &   &   &   \\ 
 \rowcolor{lightgray} Ours &  & \checkmark & \checkmark &   & 
 \checkmark & \checkmark & \checkmark &  \checkmark & \checkmark  &  \checkmark & \checkmark  & 
 \checkmark &  \checkmark &  \checkmark &  \checkmark & \checkmark  & \checkmark  & \checkmark  \\ 
\hline
\multicolumn{19}{@{}>{\footnotesize}p{1.03\linewidth}@{}}{%
\textbf{Abbreviation}: 
TML-Traditional Machine Learning, A-Attacks, D-Defenses;
AO-Attack Objective, AG-Attack Goal, AK-Attack Knowledge, 
ASt-Attack Stealthiness, ASc-Attack Scope, AI-Attack Impact, AV-Attack Variability; 
HA-Heuristic-based Attacks, LP-Label Flipping, 
FSA-Feature Space Attacks, BO-Bilevel Optimization, IM-Influence-based Method, GA-Generative Attacks.%
}
\end{tabular}
}
%\captionsetup{justification=raggedright, singlelinecheck=false}
%\caption*{\raggedright \textbf{Abbreviation}: TML-Traditional Machine Learning, A-Attacks, D-Defenses;AO-Attack Objective, AG-Attack Goal, AK-Attack Knowledge, ASt-Attack Stealthiness, ASc-Attack Scope, AI-Attack Impact, AV-Attack Variability; HA-Heuristic-based attacks, LP- Label Flipping, FSA-Feature Space Attacks, BO-Bilevel Optimization, IM-Influence-based Method, GA-Generative Attacks.}
\end{table*}

In recent years, survey studies on data poisoning attacks have steadily increased, offering a broad academic perspective on this field~\cite{pitropakis2019taxonomy,tahmasebian2020crowdsourcing,ahmed2021threats,ramirez2022poisoning}. However, most existing surveys adopt a generalized approach, encompassing a wide range of topics such as adversarial attacks and defenses across various machine learning paradigms. In particular, they often address both traditional machine learning and deep learning technologies, and frequently include discussions of other adversarial threats, such as evasion and model attacks. Although these surveys provide valuable foundational insights into data poisoning research, they lack the depth and focus needed to systematically analyze poisoning techniques tailored to deep learning models. 

With the extensive deployment of LLMs in critical applications, the risks posed by data poisoning attacks in LLMs have become increasingly concerning~\cite{alber2025medical}. Unlike conventional deep learning models, LLMs undergo multi-stage training, including pre-training, fine-tuning, and preference alignment, etc~\cite{liu2024deepseek}. Each of these stages introduces potential attack surfaces, making them uniquely susceptible to data poisoning. Despite the widespread deployment of these models, systematic research on their vulnerabilities to poisoning remains limited.

In response to these limitations, this survey focuses exclusively on data poisoning attacks within the realm of deep learning, with the goal of providing a thorough and structured analytical framework. Specifically, this work systematically reviews existing data poisoning methods in deep learning and classifies these methods across multiple dimensions. By delving into the design principles of various algorithms, this survey seeks to offer a structured reference for advancing research on data poisoning techniques in deep learning. 

Furthermore, to ensure a concentrated and thorough analysis, this survey deliberately excludes discussions on defense, with the goal of providing a more detailed and focused exploration of data poisoning algorithms themselves. Nevertheless, the systematic synthesis of attack methodologies presented in this review is expected to serve as a valuable foundation for the future development of effective defenses. 

To better highlight the distinct focus and comprehensive coverage of our work compared to prior surveys, Table I presents a comparative summary across several key dimensions. Specifically, it illustrates how existing surveys tend to cover a broad spectrum of adversarial attacks and defenses, spanning traditional machine learning and deep learning algorithms, but lack a detailed taxonomy of data poisoning techniques and algorithms. Furthermore, none of the existing surveys have systematically explored data poisoning in LLMs, despite their growing prominence in critical AI applications. In contrast, our survey offers an in-depth and systematic classification of data poisoning attacks, algorithms, and extends the discussion to LLMs, analyzing the unique attack vectors introduced by their multi-stage training paradigm.

The main contributions of this survey can be summarized as follows:

{1) Comprehensive Understanding.} This survey provides an in-depth and comprehensive overview of data poisoning in deep learning, offering a detailed theoretical framework that helps researchers better understand the mechanisms, implications, and challenges of these attacks. By focusing exclusively on data poisoning attacks, this work indirectly supports the development of more robust defense mechanisms through a thorough analysis of attack strategies.

{2) Taxonomy of Data Poisoning Attacks.} We present a systematic taxonomy of data poisoning attacks, constructed along two key dimensions to capture both conceptual and technical perspectives. The first dimension focuses on classifying attacks based on their characteristics, while the second categorizes data poisoning attacks according to the fundamental principles underlying their algorithms. By integrating these two perspectives, this survey offers a comprehensive and structured taxonomy that facilitates a clear understanding of data poisoning attacks and serves as a practical guide for researchers exploring this evolving field.

\begin{figure*}[t] 
    \centering \includegraphics[width=0.95\linewidth]{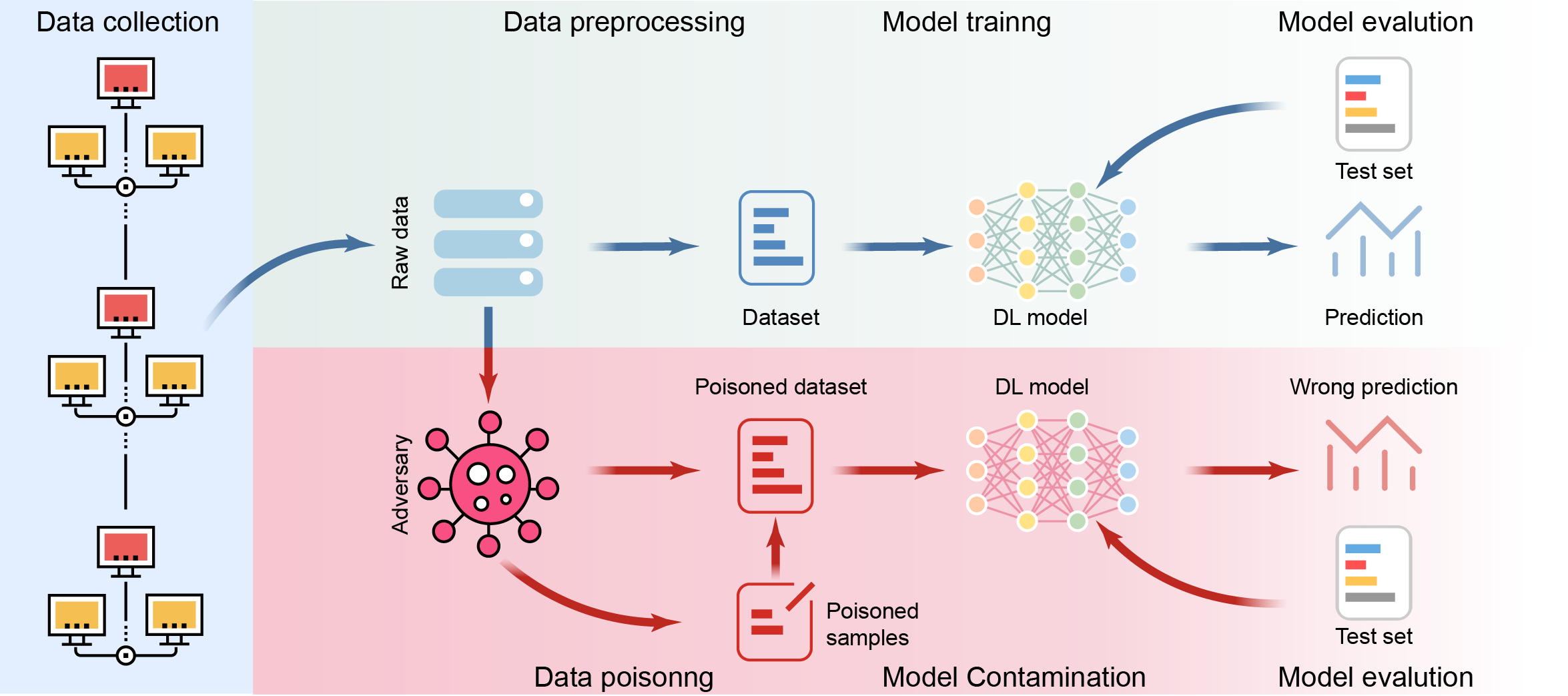}    \vspace{-0.02in}
    \caption{Deep learning and data poisoning attack pipeline.}
    \label{fig2}
     \vspace{-0.01in}
\end{figure*}

{3) Data Poisoning in Large Language Models (LLMs).} To our knowledge, this is the first survey that systematically explores data poisoning attacks in LLMs. Given the increasing deployment of LLMs in critical applications and their unique vulnerabilities across various training stages, this work provides a systematic and in-depth examination of how data poisoning can compromise these powerful models. By highlighting vulnerabilities across different training stages, we aim to fill a critical gap in the current literature.

{4) Future Research Directions.} We discuss some possible directions that could shape the future of data poisoning research. These include enhancing the stealth and effectiveness of poisoning attacks, exploring vulnerabilities in dynamic learning environments such as federated and continual learning, and expanding the study of data poisoning to multimodal AI systems and emerging architectures. Furthermore, we emphasize the need for standardized benchmarks and evaluation frameworks to ensure consistent and rigorous assessment of poisoning techniques.

{5) Online Resources and Repository.} To facilitate further research, this survey provides an open-source online repository containing a curated collection of relevant works on data poisoning attacks, including links to papers, datasets, and codes. This resource offers a convenient tool for researchers to track the latest developments and explore state-of-the-art techniques. By continuously updating the repository, this survey ensures it remains a valuable and evolving reference for the community.

The rest of this paper is organized as follows: In Section II, we introduce the fundamental concepts of data poisoning, laying the groundwork for understanding its key principles and implications. In Section III, we present the taxonomy of data poisoning attacks. In Section IV, we delve into the taxonomy of data poisoning algorithms, emphasizing the basic principles of the algorithms. Section V explores data poisoning in the context of LLMs. In Section VI, we propose potential research directions to advance the field further. Finally, we conclude this survey in Section VII.

\section{PRELIMINARY}
In this section, we provide a foundational overview of data poisoning in the context of deep learning. First, we discuss security attacks in deep learning, focusing on vulnerabilities introduced during the training phase. Next, we briefly outline the deep learning training pipeline and the fundamental concepts of data poisoning to aid in understanding and analyzing the subsequent sections of this survey. Finally, we introduce the characteristics posed by data poisoning in deep learning.

\subsection{Security Attacks in Deep Learning}
Deep learning is a cornerstone of modern artificial intelligence, enabling breakthroughs in computer vision, natural language processing, and autonomous systems~\cite{bengio2021deep}. Its success relies on large, high-quality datasets for training sophisticated models. However, as these models are increasingly deployed in high-stakes applications such as healthcare, finance, and autonomous vehicles, ensuring their reliability and robustness has become a growing concern~\cite{kumar2023impact,muley2023risk,habbal2024artificial}.

One of the most critical vulnerabilities lies in the reliance on large-scale datasets, often aggregated from diverse and potentially untrusted sources~\cite{dilmaghani2019privacy}. This dependency creates opportunities for attackers to exploit the training phase, allowing attackers to inject malicious data that undermines the model integrity before deployment. Unlike adversarial examples, which perturb inputs at the inference stage, data poisoning attacks target the training process, embedding systematic biases or backdoors that persist across deployments, compromising model reliability. These risks highlight the importance of understanding and mitigating data poisoning attacks in deep learning~\cite{chaalan2024path}.

\subsection{Data Poisoning Attack Pipeline}
Deep learning models are typically trained in a pipeline involving data collection, preprocessing, model training, and evaluation, as illustrated in Fig.~\ref{fig2}. Data poisoning refers to the deliberate injection of carefully crafted malicious samples into the training dataset of a machine learning model, with the intent of manipulating its behavior. By altering the data distribution during training, attackers can achieve a variety of objectives, ranging from degrading overall model performance to implanting specific vulnerabilities that can be exploited post-deployment.

The general framework for a data poisoning attack involves the following stages:

\begin{itemize}
    \item \textit{Data Collection.} Attackers exploit vulnerabilities in the data acquisition pipeline, such as reliance on open-source datasets, web scraping, or third-party data vendors. Poisoned data can be introduced at this stage without immediate detection.

    \item \textit{Data Poisoning.} Poisoned samples are carefully crafted to blend with benign data and evade anomaly detection mechanisms. Attackers may modify existing samples or insert synthetic data to manipulate the training distribution. 

    \item \textit{Model Contamination.} The poisoned dataset skews the learning process of the model, causing it to associate irrelevant or deceptive features with target labels, ultimately distorting decision boundaries.
    \item \textit{Evaluation and Deployment.} Once deployed, the compromised model exhibits vulnerabilities, such as targeted misclassifications or specific responses to hidden triggers.
\end{itemize}

\subsection{Characteristics of Data Poisoning in Deep Learning}
Data poisoning attacks in deep learning differ significantly from traditional machine learning poisoning attacks due to the extensive data volumes and high model complexity. The characteristics of deep learning amplify the impact and complexity of data poisoning attacks. The following are the unique characteristics of data poisoning attacks in deep learning:

The general framework for data poisoning attacks involve the following stages:

\begin{itemize}
    \item \textit{Dependence on Large-Scale Data.} Deep learning models rely on massive datasets collected from diverse and often unverified sources. The sheer volume and heterogeneity of the data make it impractical to thoroughly inspect and validate every sample, increasing the likelihood of introducing poisoned data.

    \item \textit{High Model Complexity.} Deep neural networks have substantial capacity and are capable of memorizing outliers or poisoned samples without significantly degrading performance on benign samples. This ability to generalize while retaining specific patterns makes it easier for attackers to embed malicious behavior that remains dormant until activated under specific conditions~\cite{chen2024over}.

    \item \textit{Distributed Training Environments.} Decentralized training paradigms, such as federated learning, involve multiple participants contributing data and computational resources. Malicious actors within these environments can inject poisoned data without direct access to the global model, complicating detection and accountability~\cite{bagdasaryan2020backdoor,tolpegin2020data,kasyap2024beyond}.
    \item \textit{Evaluation Difficulties.} The nonlinearity and high dimensionality of deep learning models make it difficult to assess the full impact of poisoning attacks. Existing evaluation metrics and tools are often insufficient for quantifying the subtle yet significant effects of such attacks~\cite{ozkan2024comprehensive}.
\end{itemize}

\begin{figure*}[t] 
    \centering \includegraphics[width=0.90\linewidth]{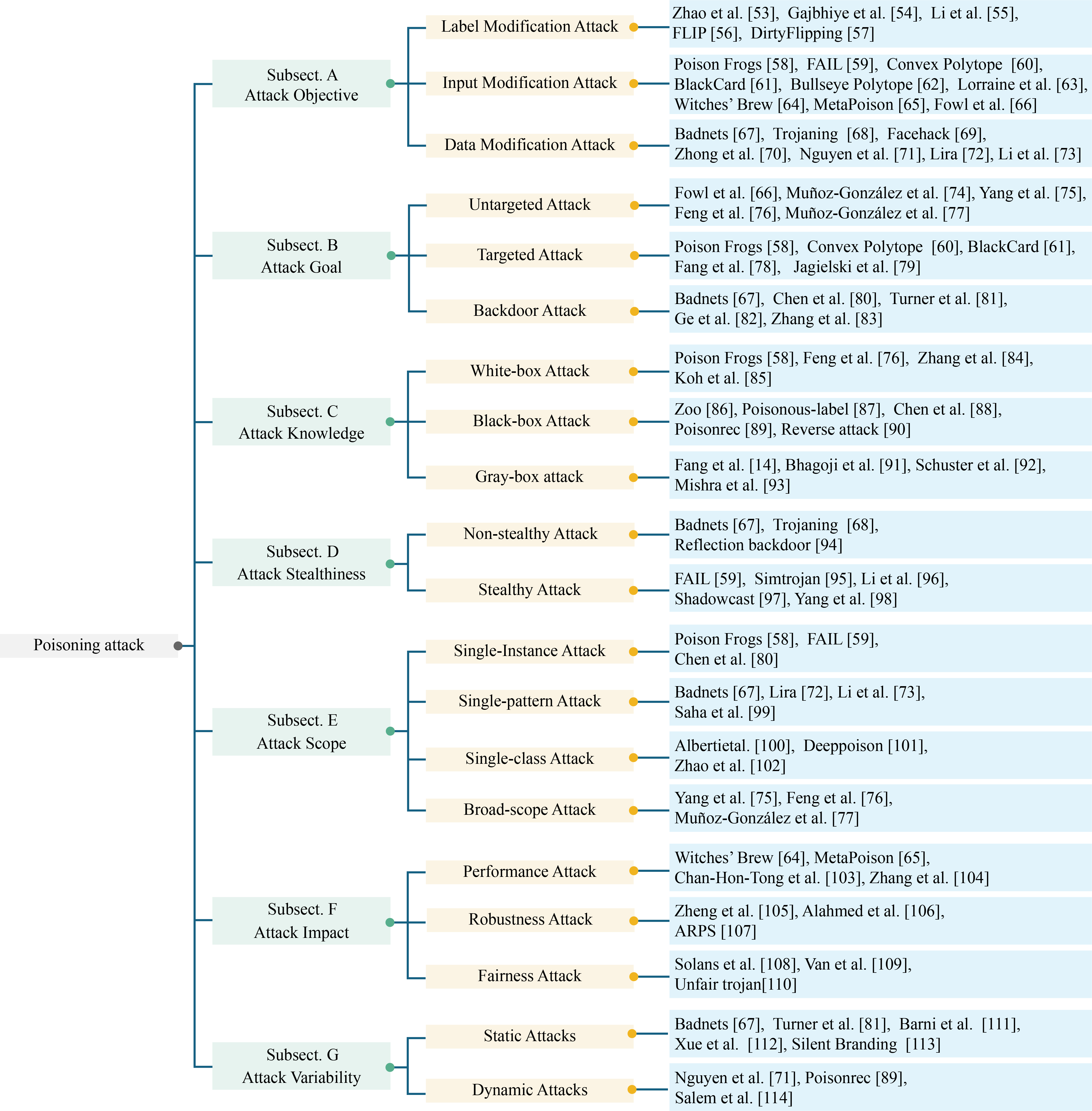}    \vspace{-0.02in}
    \caption{Taxonomy of data poisoning attacks in deep learning with representative examples.}
    \label{fig3}
     \vspace{-0.01in}
\end{figure*}

\section{TAXONOMY OF POISONING ATTACKS}
Poisoning attacks were first introduced by Barreno et al.~\cite{barreno2006can} and have since emerged as a major security threat during the training phase of machine learning systems, particularly in the context of deep learning~\cite{hu2020data}. These attacks exploit vulnerabilities in the training process by manipulating data, thereby compromising the integrity, availability, or reliability of the resulting model~\cite{bajaj2024state}. Over time, poisoning attacks have garnered significant attention due to their potential to undermine critical applications of deep learning~\cite{zhang2023generative,nguyen2024manipulating}.

In this section, we systematically categorize poisoning attacks by reviewing representative studies, as illustrated in Fig.~\ref{fig3}. The taxonomy is constructed along seven distinct dimensions, focusing on the characteristics of the attacks, including their objectives, goals, attacker knowledge, and other relevant factors. This taxonomy provides a structured framework for systematically analyzing and comparing the various characteristics of data poisoning attacks. By categorizing these attacks, this section aims to clarify their defining characteristics, differentiate attack types, and lay the groundwork for further research and the development of effective defense strategies.

\subsection{Attack Objective}
Data poisoning attacks can be categorized based on their objectives into three main types: label attack (label modification or label flipping attack)~\cite{xu2022rethinking}, input attack (input modification or clean-label attack)~\cite{goldblum2022dataset}, and data attack (data modification)~\cite{wang2022threats}. Each type targets different manipulated-target of the training data, affecting the performance of deep learning models.

\textit{Label Modification Attacks.} Label modification attacks aim to disrupt model training by altering the labels of selected samples while keeping the input features unchanged ~\cite{zhao2017efficient,gajbhiye2022data,li2022label,jha2023label,mengara2024backdoor}. A typical example is the label-flipping attack, where the labels of specific samples are deliberately changed to incorrect classes, distorting the model’s decision boundary. Zhao et al.~\cite{zhao2017efficient}, proposed a bilevel optimization-based label manipulation method, solved using a Projected Gradient Ascent (PGA) algorithm, which is both effective and transferable to black-box models. Jha et al.~\cite{jha2023label} proposed the FLIP framework, which leverages trajectory matching to perform label-only backdoor attacks.

\textit{Input Modification Attacks.} Input modification attacks, also known as clean label attacks, involve perturbing the input features of training samples while preserving their original labels. Feature collision, a type of clean-label attack, generates adversarial examples aligned with target samples in the feature space. This alignment causes the model to associate these examples with the target class during inference~\cite{shafahi2018poison,suciu2018does,zhu2019transferable,guo2020practical,aghakhani2021bullseye}. Bilevel poisoning can also retain the original labels. It employs a bilevel optimization framework to create perturbed inputs, which degrade the model's performance during training~\cite{lorraine2020optimizing,geiping2020witches,huang2020metapoison,fowl2021preventing}.

\textit{Data Modification Attacks.} Data modification attacks combine alterations to both input features and labels, or creating entirely fabricated samples to achieve the attacker’s goals. Most backdoor attacks fall under this category, where trigger patterns are embedded in the data to cause the model to exhibit specific behavior under predefined conditions~\cite{gu2017badnets,liu2018trojaning,sarkar2020facehack,zhong2020backdoor}. Generative methods, such as attacks utilizing generative adversarial networks (GANs), can produce high-quality synthetic samples that are indistinguishable from legitimate data, effectively poisoning the training process while evading detection~\cite{nguyen2020input,doan2021lira,li2021invisible}. By enabling simultaneous manipulation of input features and labels, data modification attacks offer greater potential to execute sophisticated and highly targeted attacks, significantly undermining the model’s security and reliability.

\subsection{Attack Goal}
The goal of data poisoning attacks is to manipulate the training data to compromise the performance or integrity of machine learning models. Based on the adversarial intent, these attacks are categorized into untargeted attack (indiscriminate attack or availability attack), targeted attack (integrity attacks), and backdoor attacks (integrity attacks). Below, we discuss these attack types along with relevant studies.

\textit{Untargeted Attacks.} Untargeted attacks seek to degrade the overall performance of a machine learning system, reducing its usability for legitimate tasks. These attacks often involve injecting corrupted samples or perturbing existing data to cause widespread misclassification~\cite{munoz2017towards,yang2017generative,feng2019learning,fowl2021preventing,munoz2019poisoning}. Yang et al.~\cite{yang2017generative} propose a generative framework for poisoning attacks on neural networks, using an autoencoder to efficiently create poisoned samples. Reference~\cite{feng2019learning}  describes an approach where the attacker utilizes a generative model to produce clean-label poisoning samples designed to undermine the victim model. Fowl et al.~\cite{fowl2021preventing} employed a gradient alignment optimization method to subtly modify the training data, effectively reducing the model's performance. These studies illustrate the variety of techniques attackers can leverage to compromise the reliability and overall functionality of deep learning models.

\textit{Targeted Attacks.} Targeted attacks aim to mislead the model on specific target samples while preserving its general performance on clean data. Such targeted attacks are particularly insidious, as they ensure the system appears functional to most users~\cite{shafahi2018poison,zhu2019transferable,
fang2020influence,guo2020practical,jagielski2021subpopulation}. Shafahi et al.~\cite{shafahi2018poison} proposed a poisoning framework for deep learning models that generates subtle perturbations to achieve targeted misclassification. Jagielski et al.~\cite{jagielski2021subpopulation} proposed a subpopulation attack that targets the performance of a machine learning model on a specific subpopulation, while preserving its accuracy on the remaining dataset.

\textit{Backdoor Attacks.} Backdoor attacks involve embedding a trigger pattern into the training data, enabling the adversary to manipulate model predictions when the trigger is present, while maintaining high accuracy on clean samples~\cite{gu2017badnets,chen2017targeted,turner2019label,ge2023data,zhang2024data}. Gu et al.~\cite{gu2017badnets} introduced ``BadNets", which demonstrated that injecting a small number of poisoned samples with an embedded trigger into the training data can effectively implant a backdoor into the model. Chen et al.~\cite{chen2017targeted} advanced backdoor attacks by introducing imperceptible noise patterns as triggers, making detection even more challenging. Another variation, proposed by Turner et al.~\cite{turner2019label}, utilizes label-consistent backdoor triggers to minimize visual discrepancies between poisoned and clean samples, further evading detection. Backdoor attacks demonstrate the severe risks posed by data poisoning in outsourced or collaborative deep learning workflows.

\subsection{Attack Knowledge}
The attacker's knowledge plays a critical role in determining the feasibility and effectiveness of data poisoning attacks. Based on the level of access to the victim's machine learning system, attacks can be categorized into three types: Black-box attacks, Grey-box attacks, and White-box attacks. This classification reflects the attacker’s access to information about the model architecture, parameters, and training data. Understanding these categories is crucial for evaluating the feasibility and impact of different attack strategies in practical scenarios.

\textit{White-box attacks.} In this setting, the adversary has full knowledge of the victim model, including its architecture, parameters, and the training dataset. This extensive knowledge allows the adversary to craft highly precise poisoning samples that maximize the attack's effectiveness. Techniques such as bilevel optimization are often employed in white-box scenarios to identify and inject poisoned samples that directly manipulate the model's decision boundaries. While white-box attacks are not always feasible in real-world applications due to the difficulty of obtaining such privileged access, they serve as an important benchmark for evaluating a worst-case~\cite{shafahi2018poison,feng2019learning,zhang2019poisoning,koh2022stronger}.

\textit{Black-box Attacks.} In this case, the attacker has no direct access to the victim model's internal details, such as its architecture, parameters, or training data. Instead, the adversary interacts with the system through its input-output behavior, typically by querying the model and observing its predictions. These attacks rely on techniques such as transfer learning, where a surrogate model is used to approximate the target model’s behavior. Black-box poisoning attacks are particularly challenging to detect, as they exploit limited information while still effectively degrading the model’s performance~\cite{chen2017zoo,liu2021poisonous,chen2023black}. For instance, black-box attacks on recommender systems manipulate user-item interaction data without requiring access to the underlying recommendation algorithm~\cite{song2020poisonrec,zhang2021reverse}.

\textit{Grey-box Attacks.} The attacker has partial knowledge of the victim model, such as the architecture or a subset of the training data, but lacks full access to critical details like the complete training set or exact parameter values. This intermediate level of knowledge enables the attacker to perform more targeted and efficient poisoning compared to black-box scenarios~\cite{bhagoji2019analyzing,fang2020local,schuster2021you,mishra2023towards}. For example, an adversary with access to a pre-trained model but not the fine-tuning data may craft poisoning samples to disrupt the fine-tuning process. Grey-box attacks highlight the risks associated with shared or collaborative training environments, where some information about the model may be inadvertently exposed.

\subsection{Attack Stealthiness}
Based on the detectability of adversarial modifications in the training data, poisoning attacks can be classified into non-stealthy attack and stealthy attack.

\textit{Non-stealthy Attacks.} Non-stealthy attacks involve noticeable modifications to the training data, often injecting conspicuous anomalies or significantly altering existing samples. These attacks prioritize maximizing their impact on the model's performance rather than evading detection. For instance, in~\cite{gu2017badnets,liu2018trojaning,liu2020reflection}, the authors introduced trojaning attacks on neural networks, where an attacker embeds malicious behavior by reverse-engineering the model to generate a trigger and retraining it with synthetic data. The trigger activates specific neurons, causing the model to produce adversarial outputs while maintaining normal performance on benign inputs. 

\textit{Stealthy Attacks.} In contrast to non-stealthy attacks, stealthy attacks leverage subtle modifications to the training data, ensuring the poisoned samples remain statistically similar to clean data and thus evade detection. These attacks are particularly challenging to defend against as they carefully balance perturbation magnitude and attack efficacy~\cite{ren2021simtrojan,suciu2018does,li2020invisible,xu2024shadowcast,yang2024stealthy}. Suciu et aly~\cite{suciu2018does} introduced the "StingRay" attack, which leverages targeted poisoning to alter the decision boundary of machine learning models. The StingRay attack generates poisoned samples that are highly similar to clean training instances in the feature space. Li et al.~\cite{li2020invisible} introduced two novel techniques for embedding triggers into models: steganography and regularization. These methods effectively ensure the success of the attack while preserving the original functionality of the model and achieving a high level of stealth. Xu et al.~\cite{xu2024shadowcast} introduced ``Shadowcast," a stealthy data poisoning attack specifically targeting vision-language models (VLMs). Shadowcast constructs visually indistinguishable poisoned image-text pairs, manipulating the model to generate adversarial outputs while maintaining coherence and subtlety.

\subsection{Attack Scope}
Poisoning attacks can be categorized by the breadth of their impact on the target model, ranging from narrowly focused attacks on individual samples to broader attacks that disrupt multiple categories or entire datasets. These classifications are useful for understanding the scalability and specificity of the threat posed by poisoning attacks.

\textit{Single-instance Attacks.} Single-instance attacks focus on inducing misclassification of a specific target sample without affecting the overall model performance~\cite{chen2017targeted,shafahi2018poison,suciu2018does}. For instance, the feature collision method introduced by Shafahi et al. generates adversarial examples in the training set that cause a specific test instance to be misclassified into a desired class~\cite{shafahi2018poison}. These attacks often require precise feature manipulation and are commonly applied in clean-label settings to remain undetectable.

\textit{Single-pattern attacks.} Single-pattern attacks aim to misclassify a group of inputs sharing a specific pattern. The backdoor attack strategy is a prominent example, where attackers introduce a trigger pattern during training that reliably activates a specific malicious behavior at inference~\cite{gu2017badnets,saha2020hidden,doan2021lira,li2021invisible}. These attacks demonstrate high stealth and flexibility, as the triggers are often imperceptible yet effective in subverting the model.

\textit{Single-class Attacks.} In single-class attacks, the objective is to disrupt the classification of all instances belonging to a particular class while preserving the performance of the model on other classes. Methods such as bilevel optimization have been shown to craft poison samples that compromise an entire class with minimal perturbations~\cite{alberti2018you,chen2021deeppoison,zhao2022towards}. This attack type is particularly concerning in scenarios involving sensitive classifications, such as biometric systems or medical diagnostics.

\textit{Broad-scope Attacks.} Broad-scope attacks seek to degrade the performance of the model across multiple classes or the entire dataset. These attacks are often employed in availability attacks, where the goal is to render the model unusable or significantly degrade its general performance~\cite{yang2017generative,munoz2019poisoning,feng2019learning}. For instance, attackers might inject a large number of poisoned samples into the training set, affecting the model's overall accuracy and reliability, thus undermining its ability to function correctly in a real-world environment.

\subsection{Attack Impact}

Poisoning attacks can be categorized according to the specific impact they are designed to achieve, including performance attack, robustness attack, and fairness attack. Each category addresses a different dimension of vulnerability exploited by attackers.

\textit{Performance Attacks.} The primarily of this type of attack is to degrade the overall accuracy or usability of the model by introducing poisoned samples into the training dataset. These attacks often exploit weaknesses in the training process to induce widespread misclassification. For example, methods such as bilevel optimization and reinforcement learning have been used to craft poisoning samples that disrupt the overall performance of models across tasks like image classification systems~\cite{chan2018algorithm,huang2020metapoison,geiping2020witches,zhang2022backdoor}.

\textit{Robustness Attacks.} Instead of targeting overall performance, robustness attacks focus on undermining the resilience of the model to perturbations or adversarial examples, making it more vulnerable to malicious inputs during deployment. Zheng et al.~\cite{zheng2022concealed} proposed a concealed poisoning attack that reduces the robustness of deep neural networks by generating poisoned samples through a bi-level optimization framework. This approach not only degrades the model's resistance to adversarial inputs but also ensures high stealth by maintaining performance on clean samples. Similarly, Alahmed et al.~\cite{alahmed2024impacting} investigated the impact of poisoning attacks on deep learning-based network intrusion detection systems. \cite{jiang2024adversarial} proposes an Adversarial Robustness Poisoning Scheme (ARPS) that subtly degrades a model's adversarial robustness while preserving its normal performance, posing a stealthy threat to deep neural networks.

\textit{Fairness Attacks.} Fairness attacks target the ethical and equitable functioning of the model by skewing its decisions toward biased outcomes. These attacks exploit vulnerabilities in the training process to introduce systemic biases. For instance, solans et al.~\cite{solans2020poisoning} proposed a gradient-based poisoning attack designed to increase demographic disparities among groups by introducing classification inequities. This approach effectively manipulates model behavior in both white-box and black-box scenarios, showcasing the adaptability of such attacks across different settings. Similarly, Van et al.~\cite{van2022poisoning} developed a framework for generating poisoning samples through adversarial sampling, labeling, and feature modification. This framework enables attackers to adjust their focus on fairness violation or accuracy degradation, demonstrating significant impacts on group-based fairness notions such as demographic parity and equalized odds. Furth et al.~\cite{furth2024unfair} introduced the ``Un-Fair Trojan" attack, a backdoor approach that targets model fairness while remaining highly stealthy. This method uses a trojan trigger to disrupt the fairness metrics of the model, significantly increasing demographic parity violations without reducing overall accuracy.

\subsection{Attack Variability}
The persistence and adaptability of data poisoning attacks can vary significantly depending on how the adversary injects and maintains poisoned data within the system. While some attacks rely on fixed, predefined manipulations, others continuously evolve, making detection and mitigation more challenging. This distinction leads to two primary categories: static attacks and dynamic attacks. 

\textit{Static Attacks.} Static attacks involve injecting poisoned data into the training set before the model is trained, with the adversarial modifications remaining unchanged throughout the entire lifecycle of the model. These attacks often use fixed persistent poisoning strategies that do not adapt post-deployment. Most data poisoning attacks, including label-flipping and fixed-pattern backdoor attacks, fall under this category~\cite{turner2019label,barni2019new,xue2021backdoors,jang2025silent}. A well-known example of static poisoning is BadNets~\cite{gu2017badnets}, which embeds a fixed trigger into training samples, causing the model to consistently misclassify inputs containing the trigger at inference time. While static attacks can be highly effective, their reliance on fixed patterns makes them more susceptible to detection by anomaly detection and robust training techniques.

\textit{Dynamic attacks.} Dynamic attacks introduce adaptive poisoning strategies, where the attack evolves over time to enhance stealth and robustness. Salem et al.~\cite{salem2020dynamic} introduced dynamic backdoor attacks, such as Backdoor Generating Network (BaN) and Conditional BaN (c-BaN), which generate variable triggers across different inputs. These adaptive triggers allow backdoored models to evade traditional defenses like Neural Cleanse and STRIP. Similarly, Nguyen et al.~\cite{nguyen2020input} proposed input-aware backdoor attacks, where triggers are generated conditionally based on the input data, making detection significantly more challenging. Beyond backdoor attacks, PoisonRec by Song et al.~\cite{song2020poisonrec} presents an adaptive poisoning framework for black-box recommender systems, using reinforcement learning to iteratively refine poisoning strategies based on system feedback. This method demonstrates that adaptive poisoning can extend beyond classification models to real-world recommender systems, increasing the persistence and effectiveness of attacks.

\begin{table*}[htbp]
\centering
\caption{Summary of data poisoning algorithms}
\resizebox{\textwidth}{!}{
\footnotesize
\renewcommand{\arraystretch}{1.1} % 增加行距，使表格更美观

\begin{tabular}{>{\centering\arraybackslash}m{3cm} >{\centering\arraybackslash}m{2cm} >{\centering\arraybackslash}m{0.7cm} >{\centering\arraybackslash}m{0.7cm} >{\centering\arraybackslash}m{3cm} >{\centering\arraybackslash}m{3cm}}
\hline
\textcolor{headercolor}{Algorithms} & \textcolor{headercolor}{Attacks} & \textcolor{headercolor}{Ref.} &\textcolor{headercolor}{Year} &\textcolor{headercolor}{Model} &\textcolor{headercolor}{ Application}\\
\hline
\rowcolor{lightgray} 
{Heuristic-based Attacks} 
& BadNets & \cite{gu2017badnets} & 2017 & CNN, Faster-RCNN & handwritten digit recognition, traffic sign detection\\
\rowcolor{lightgray}
&Chen et al.&\cite{chen2017targeted} &2017 & DeepID, VGG-Face  &face recognition,  face verification\\
\rowcolor{lightgray}
& Alberti et al.&\cite{alberti2018you}&2018 & lexNet, VGG-16, ResNet-18, DenseNet-121  & image classification\\
\rowcolor{lightgray}
& Li et al. & \cite{li2021invisible}& 2021 & ResNet-18 & image classification\\
\rowcolor{lightgray}
& Pixdoor & \cite{arshad2021pixdoor}& 2021 & LeNet  & handwritten digit recognition\\

{Label Flipping Attacks} 
& Zhang et al. & \cite{zhang2021understanding} & 2021 & AlexNet, Inception V3 & image classification \\
& Zhang et al. & \cite{zhang2021label} & 2021 & AlexNet, LeNet & spam classification \\
& Li et al. & \cite{li2022label} & 2022 & MLP & malware detection \\
& FLIP & \cite{jha2023label} & 2023 & ResNet-32, ResNet-18, VGG-19, Vision Transformer & image classification \\
&  Lingam et al. & \cite{lingam2023rethinking} & 2024 & GCN, GAT, APPNP, CPGCN, RTGNN & node classification. \\

\rowcolor{lightgray} 
{Feature Space Attacks} 
& Poison Frogs & \cite{shafahi2018poison} & 2018 & InceptionV3, AlexNet & image classification\\
\rowcolor{lightgray}
& FAIL & \cite{suciu2018does} & 2018 & NN & malware detection, image classification, exploit prediction, data breach prediction\\
\rowcolor{lightgray}
& Convex Polytope & \cite{zhu2019transferable} & 2019 & ResNet-18, ResNet-50, DenseNet-121, SENet-18 & image classification\\
\rowcolor{lightgray}
& Saha et al. & \cite{saha2020hidden} & 2020 & RAlexNet & image classification\\
\rowcolor{lightgray}
& BlackCard & \cite{guo2020practical} & 2020 & ResNet, DenseNet  & image classification, face recognition\\
\rowcolor{lightgray}
& Bullseye Polytope & \cite{aghakhani2021bullseye} & 2021 & SENet18, ResNet50, ResNeXt29-2x64d, DPN92, MobileNetV2, GoogLeNet  & image classification\\
\rowcolor{lightgray}
& Luo et al. & \cite{luo2022enhancing} & 2022 & ResNet18  & image classification\\

{Bilevel Optimization Attacks} 
& Mu\~{n}oz-Gonz\'{a}lez et al. & \cite{munoz2017towards} & 2017 & CNN & spam filtering, malware detection, handwritten digit recognition \\
& MetaPoison & \cite{huang2020metapoison} & 2020 & ConvNetBN, VGG13,ResNet20 & image classification \\
& Witches' Brew & \cite{geiping2020witches} & 2020 & ConvNet, ResNet-18 & image classification \\
& Pourkeshavarz et al. & \cite{pourkeshavarz2024adversarial} & 2024 & PGP, LaPred, HiVT, TNT, MMTransformer, LaneGCN & trajectory prediction \\
& BLTO & \cite{sun2024backdoor} & 2024 & SimCLR, BYOL, SimSiam & feature extractor \\

\rowcolor{lightgray} 
{Influence-based Attacks} 
& Koh and Liang & \cite{koh2017understanding} & 2017 & CNN &  understanding model behavior, debugging models, detecting dataset errors\\
\rowcolor{lightgray} 
& Basu et al. & \cite{basu2020influence} & 2021 & small CNN, LeNet , ResNets, VGGNets & image classification\\
\rowcolor{lightgray} 
& Koh et al. & \cite{koh2022stronger} & 2022 & neural network & spam detection, sentiment classification\\

{Generative Attacks} 
& Yang et al. & \cite{yang2017generative} & 2017 & auto-encoder & image classification\\
& Feng et al. & \cite{feng2019learning} & 2019 & auto-encoder & image classification\\
& Mu\~{n}oz-Gonz\'{a}lez et al.& \cite{munoz2019poisoning} & 2019 & pGAN & image classification\\
& Psychogyios et al.& \cite{psychogyios2023gan} & 2023 & GAN & grapevine image classificationn\\
& Chen et al.& \cite{chen2024poisoning} & 2024 & GAN & cloud API recommender\\

\hline
\end{tabular}
}
\end{table*}

\begin{figure*}[t] 
    \centering \includegraphics[width=0.95\linewidth]{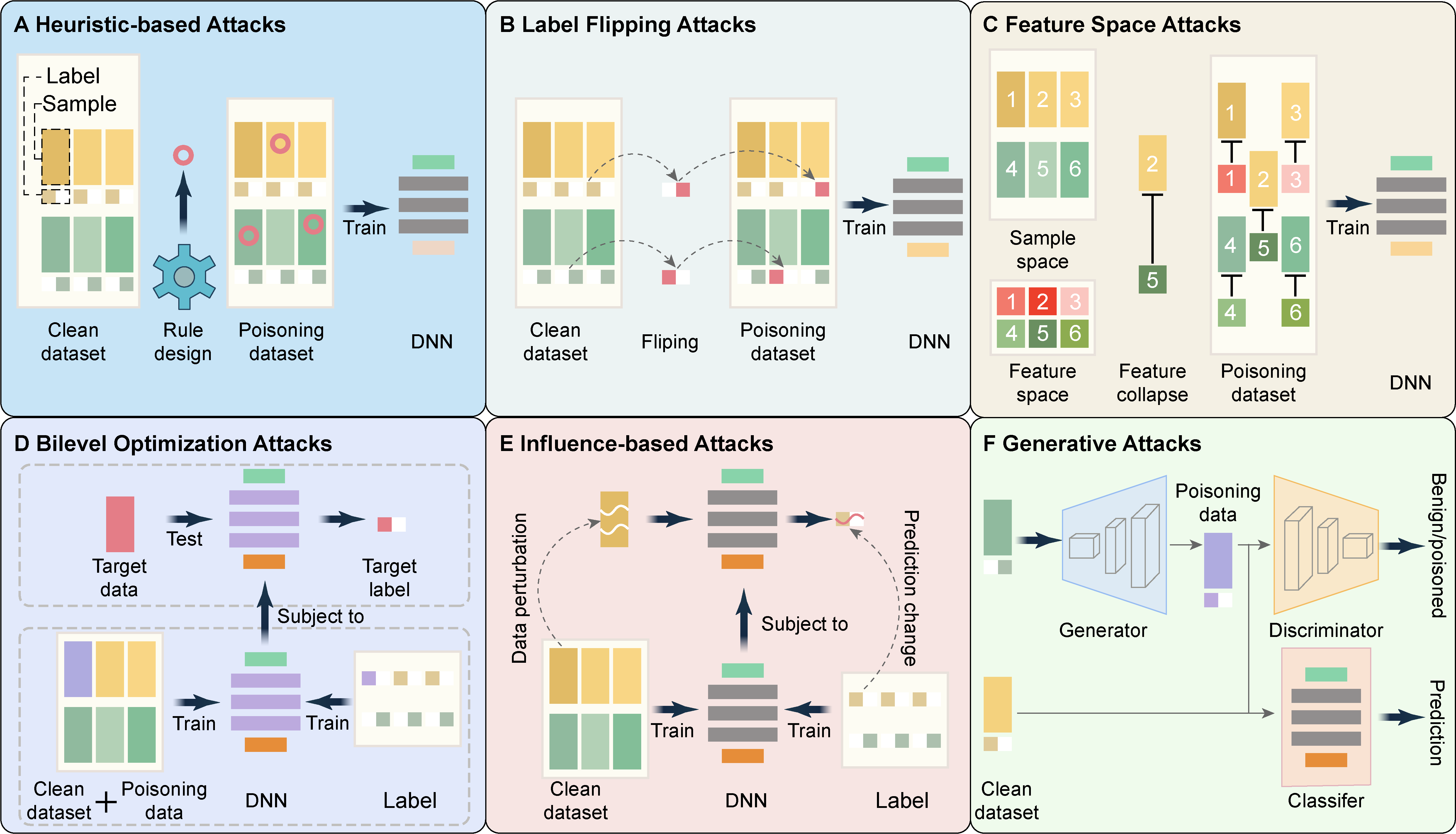}    \vspace{-0.02in}
    \caption{The sub-taxonomy of data poisoning algorithms.}
    \label{fig4}
     \vspace{-0.01in}
\end{figure*}

\section{DATA POISONING ALGORITHMS}

The effectiveness of data poisoning attacks is determined by the strategies used to manipulate training data. These strategies differ in terms of computational complexity, the extent of adversarial control, and their ability to evade detection. A precise understanding of these techniques is crucial for both the development of new poisoning methods and the design of effective defenses.

This section examines the fundamental principles and mathematical foundations of poisoning algorithms, focusing on how they alter training data, optimize attack objectives, and exploit model vulnerabilities. Fig.~\ref{fig4} illustrates six major algorithmic approaches used in data poisoning: heuristic-based attacks, label flipping, feature collision attacks, bilevel optimization, influence-based methods, and generative model-based attacks.  To provide a comprehensive overview, Table II presents a summary of data poisoning algorithms. The following subsections provide a detailed analysis of these methods:

\subsection{Heuristic-based Attacks}
Heuristic-based attacks represent one of the fundamental approaches to data poisoning in deep learning. These attacks leverage predefined heuristic rules rather than complex optimization frameworks to craft poisoned data samples. Unlike sophisticated bilevel optimization or influence-based methods, heuristic-based approaches typically rely on empirical observations and domain knowledge to manipulate training data. While they are often less effective in achieving highly targeted attacks, their simplicity and efficiency make them widely applicable, especially in scenarios where computational constraints exist.

A foundational heuristic-based attack is BadNets~\cite{gu2017badnets}, which introduces a straightforward poisoning strategy: a small trigger pattern is injected into selected training samples, and their labels are modified to a target class. This results in a trained model that behaves normally on clean inputs but misclassifies any input containing the trigger. However, BadNets-style attacks often introduce detectable data anomalies, making them susceptible to defensive techniques such as anomaly detection and adversarial training.

To enhance stealth, subsequent research introduced more sophisticated injection techniques. Chen et al~\cite{chen2017targeted} demonstrated that only a small number of poisoned samples without direct access to the model are sufficient to implant an effective backdoor. Similarly, Alberti et al.~\cite{alberti2018you} explored a minimal perturbation approach, showing that modifying just a single pixel per image across the dataset can establish a functional backdoor. This attack emphasizes how seemingly negligible changes can have disproportionate effects on deep learning models. An even more covert strategy was proposed in~\cite{li2020invisible}, which improves upon BadNets by embedding triggers in a way that remains imperceptible to both human inspectors and anomaly detection algorithms. This is achieved through steganographic techniques, ensuring that backdoors remain hidden while maintaining a high attack success rate. Pixdoor~\cite{arshad2021pixdoor} created a nearly undetectable backdoor attack with minimal poisoned data injection by manipulating only the least significant bits of pixel values.

\subsection{Label Flipping Attacks}

High-quality training data is essential for ensuring the accuracy and reliability of machine learning models. Since these models learn by associating input data with correct labels, any disruption in this relationship weakens their performance. Label flipping attacks exploit this vulnerability by selectively altering training labels while leaving the actual data unchanged. While these attacks do not maintain clean labels, they avoid introducing visible artifacts that could make manipulation obvious. By embedding false associations into the learning process, these attacks mislead the model, causing systematic errors or targeted misclassification. A  label flipping attack can be formally expressed as follows:
\begin{equation} 
LF(y) = \left\{
\begin{array}{ll}
1 - y, & y \in Y = \{0, 1\} \\
\text{random}(Y \setminus \{y\}), & \text{others}
\end{array}
\right.
\end{equation}
where $y$ represents the original label, and $1 - y$ denotes the flipped label in a 0/1 classification task. For multi-class classification, the function $random(\cdot)$ selects samples randomly for label flipping.

In addition to randomly selecting, an attacker can strategically choose a subset of data for flipping to achieve a stronger attack impact. To understand how label flipping affects deep learning models, Zhang et al.~\cite{zhang2021understanding} conducted experiments to evaluate the impact of random label corruption on neural networks. Their study revealed that deep models can perfectly fit training data even when labels are completely randomized, achieving zero training error while failing to generalize on test data. This overfitting behavior demonstrates that deep learning models do not inherently distinguish between correct and incorrect labels, making them highly susceptible to label flipping attacks. Furthermore, standard regularization techniques such as weight decay and dropout were found to be ineffective in mitigating the effects of mislabeled data. Since models trained on flipped labels internalize incorrect associations, their decision boundaries become distorted, leading to long-term degradation in performance. While victims may detect label flipping through abnormal test errors, by the time this occurs, significant computational resources have already been expended, making it an effective denial-of-service attack.

Label flipping attacks have also been explored in different domains and model architectures, revealing their wide-ranging impact. Zhang et al.~\cite{zhang2021label} demonstrated that label flipping is particularly effective in spam filtering systems, where flipping a small fraction of spam labels to legitimate ones significantly degrades detection accuracy. This highlights the broader risk of label flipping in security-sensitive applications where models rely on clean labels for decision-making. Further, Li et al.~\cite{li2022label} proposed a targeted label flipping strategy using clustering techniques to identify and alter the most influential samples. Their method showed that strategically flipping a small subset of labels can be far more damaging than random flipping, making detection and mitigation more challenging.

Beyond degrading overall model performance, label flipping can also be used for backdoor injection. Jha et al.~\cite{jha2023label} introduced FLIP, a label-only backdoor attack demonstrating that attackers can implant backdoors in models without modifying input features. Their study revealed that even corrupting just 2\% of labels in datasets like CIFAR-10 could achieve near-perfect backdoor success rates, posing a major risk in crowd-sourced training environments. This vulnerability is even more pronounced in graph neural networks (GNNs), as shown by Lingam et al.~\cite{lingam2023rethinking}, who found that flipping even a single label in graph-based models can significantly disrupt classification accuracy due to the structural dependencies between data points.

\subsection{Feature Space Attacks}

Feature space attacks manipulate training data so that the feature representations of poisoned samples become indistinguishable from a specific target. Unlike label flipping poisoning attacks that alter labels, this method focuses on feature space manipulation, ensuring that poisoned samples appear natural while influencing the decision boundary of the model. By injecting carefully crafted samples into the training set, the attacker can make a chosen target sample misclassified without modifying the target sample itself. Feature space poisoning attacks offer three key stealth advantages. First, modifying feature space associations does not require altering labels, making the attack highly inconspicuous. Second, the attack can introduce only minor perturbations to input samples using optimization-based techniques, without embedding noticeable poisoning patterns, thereby evading manual inspection. Third, these attacks typically affect only the classification of specific target samples while leaving non-target samples unaffected, making detection significantly more challenging.

The first feature space attack was processed by Shafahi et al. in 2018~\cite{shafahi2018poison}, called feature collision attack. This attack focuses on misclassifying specific target samples while keeping the poisoned samples visually inconspicuous. This attack modifies the deep features of certain training data, making them closer to the target class in feature space, thereby misleading the model into misclassifying the target sample during inference. Its optimization objective includes maximizing the feature similarity between poisoned samples and the target class while maintaining visual similarity to the original class to enhance stealth. The optimization objective of feature collision attack is defined as follows:
\begin{equation} 
x_p = \arg\min_{x_p} \|f(x_p) - f(x_t)\|_2^2 + \beta \|x_p - x_b\|_2^2
\label{eq:FSA1}
\end{equation}
where $x_p$ is the poisoned sample, $x_t$ is the target test sample, $x_b$ is a base class sample in the training data, $f$ is the target model, $f(\cdot)$ is the output of the model, and $\beta$ is a hyperparameter. In Equation~\eqref{eq:FSA1}, the first term makes the poisoned sample close to the attack target category $t$, achieving the attack purpose. The second term controls the poisoned data to be similar to the base class data, so that there is no obvious visual difference between the two. This attack is primarily used in transfer learning, especially in scenarios where pre-trained models are fine-tuned. Poisoning only a small number of samples can effectively manipulate classification results. However, the method relies on the attacker having full knowledge of the targeted model, which limits its practicality. Additionally, if the victim model is later retrained with new clean data using end-to-end training or layer-wise fine-tuning, the effects of the poisoned data will gradually wear off.

While the feature collision attack is effective, it requires white-box access to the victim model. To overcome this limitation, Suciu et al.~\cite{suciu2018does} proposed the FAIL adversary model and introduced StingRay, a clean-label poisoning method that generalizes across multiple neural network architectures. StingRay avoids precise feature-space manipulation, improving attack transferability and stealth, making it more practical and broadly applicable.

Building on these findings, Zhu et al.~\cite{zhu2019transferable} proposed the Convex Polytope Attack, which enhances transferability by placing multiple poisoned samples around the target in feature space, increasing misclassification across different classifiers. Its optimization objective is formulated as:
\begin{equation}
\begin{aligned}
   \min_{\{c^{(i)}_j\}, \{x^{(j)}_p\}} &\frac{1}{2m} \sum_{i=1}^{m} 
    \frac{\left\| f^{(i)}(x_t) - \sum_{j=1}^{k} c^{(i)}_j f^{(i)}(x_p^{(j)}) \right\|^2}
    {\left\| f^{(i)}(x_t) \right\|^2} \\
    \text{s.t.} &\quad \sum_{j=1}^{k} c^{(i)}_j = 1, \quad c^{(i)}_j \geq 0, \quad \forall i, j; \\ 
    &\left\| x_p^{(j)} - x_b^{(j)} \right\|_{\infty} \leq \epsilon, \quad \forall j.
\end{aligned}
\label{eq:FSA2}
\end{equation}
where $x_p$ represents the poisoned sample, $x_t$  denotes the target test sample, and $x_b$ is a base-class sample from the training data. A set of pretrained models is defined as $f^{(i)}(x_t)$, with $m$ representing the number of sets in the model. The attack constructs $k$ poisoned samples, which collectively form an enclosing structure around the target in feature space. A constraint $\sum_{j=1}^{k} c_{j}^{(i)} = 1, c^{(i)}_j \geq 0$  ensures that the weights assigned to these enclosing poisoned samples are all positive and sum to one. Additionally, the upper bound of perturbation is defined as $\epsilon$, controlling the maximum modification applied to each poisoned sample. The experiments showed that Convex Polytope significantly outperforms standard feature collision attacks in black-box settings, achieving a 50\% attack success rate while poisoning only 1\% of the training data. 

Despite its strong transferability, the Convex Polytope Attack suffered from high computational costs. To address this, Bullseye Polytope~\cite{aghakhani2021bullseye} optimized the attack by centering the target within the poisoned polytope, improving stability, efficiency, and attack success rates. Meanwhile, BlackCard~\cite{guo2020practical} removed the need for model knowledge by crafting universal poisoned samples that generalize across architectures, significantly enhancing stealth. Feature-space poisoning has also been adapted for backdoor attacks, as seen in Hidden Trigger Backdoor Attacks~\cite{saha2020hidden}, which conceal triggers during training, and Luo et al.~\cite{luo2022enhancing}, who introduced image-specific triggers to further evade detection.

\subsection{Bilevel Optimization Attacks}
Bilevel optimization formalizes data poisoning attacks as a two-level problem: Inner optimization: the victim model is trained on a dataset that includes poisoned samples. Outer optimization: the attacker optimizes the poisoned data to maximize the desired attack effect, such as misclassification or performance degradation. Mathematically, bilevel optimization poisoning can be expressed as:
\begin{equation}
\begin{aligned}
& D_p' = \arg\max_{D_p} \mathcal{F}(D_p, \theta') = \mathcal{L}_{\text{out}}(D_{\text{val}}, \theta')  \\ 
& \text{s.t.} \quad \theta' = \arg\min_{\theta} \mathcal{L}_{\text{in}}(D \cup D_p, \theta)
\end{aligned}
\label{eq:BO1}
\end{equation}
where $D$, $D_{\text{val}}$, and $D_p$ represent the original training dataset, the validation dataset, and the poisoned dataset, respectively. The inner and outer loss functions are denoted as $\mathcal{L}_{\text{in}}$ and  $\mathcal{L}_{\text{out}}$. The objective of the outer optimization is to generate a poisoned dataset that maximizes the classification error of the target model $\theta'$ on the clean validation dataset $D_{\text{val}}$. Meanwhile, the inner optimization iteratively updates the target model using the poisoned dataset $D \cup D_p$, ensuring that the model is trained on compromised data. Since the model parameters $\theta'$ are implicitly determined by the poisoned dataset $D_p$, the function $\mathcal{F}$ is introduced to represent the dependency between $\theta'$ and $D_p$ in the outer optimization step. The bilevel optimization process operates as follows: once the inner optimization reaches a local minimum, the outer optimization updates the poisoned dataset $D_p$ using the newly trained target model $\theta'$. This process repeats until the outer loss function $\mathcal{L}_{\text{out}}(D_{\text{val}}, \theta')$ converges. 

The above bilevel optimization attack is an indiscriminate attack, as its primary objective is to maximize the overall classification error of the target model. However, bilevel optimization can be adapted for other attack types, such as targeted attacks and backdoor attacks. In the case of a targeted attack, the bilevel problem transforms into a min-min optimization problem, formally defined as follows:
\begin{equation}
\begin{aligned}
& D_p' = \arg\max_{D_p} \mathcal{F}(D_p, \theta') = \mathcal{L}_{\text{out}}(\{x_t, y_{\text{adv}}\}, \theta')  \\ 
& \text{s.t.} \quad \theta' = \arg\min_{\theta} \mathcal{L}_{\text{in}}(D \cup D_p, \theta)
\end{aligned}
\label{eq:BO2}
\end{equation}
here, $y_{\text{adv}}$ is the incorrect target class predefined by the attacker. In this case, the objective of the outer optimization is to generate a poisoned dataset that minimizes the classification error of the target model $\theta'$ on the designated target samples, ensuring that they are misclassified into $y_{\text{adv}}$. Gradient-based approaches approximate this optimization iteratively, using the chain rule to calculate gradients if  $\mathcal{L}_{\text{out}}$ is differentiable:
\begin{equation}
\begin{aligned}
& \nabla_{D_p} \mathcal{F} = \nabla_{D_p} \mathcal{L}_{\text{out}} 
+ \frac{\partial \theta}{\partial D_p}^{\top} \nabla_{\theta} \mathcal{L}_{\text{out}}
\\
& \text{s.t.} \quad 
\frac{\partial \theta}{\partial D_p}^{\top} = (\nabla_{D_p} \nabla_{\theta} \mathcal{L}_{\text{in}}) (\nabla_{\theta}^2 \mathcal{L}_2)^{-1}
\end{aligned}
\label{eq:BO3}
\end{equation}
where $\nabla_{D_p} \mathcal{F}$ represents the partial derivative of $\mathcal{F}$ with respect to $D_p$. The poisoned data $D_p^{(i)}$ of the i-th iteration can be updated to $D_p^{(i+1)}$ by gradient ascent.

To adapt bilevel optimization to deep neural networks, researchers introduced back-gradient descent, a technique that allows attackers to approximate bilevel optimization solutions without explicitly solving the inner problem. Mu\~{n}oz-Gonz\'{a}lez et al.~\cite{munoz2017towards} first applied this approach to neural networks in 2017, demonstrating that gradient-based optimization could successfully craft poisoned data that influences deep model behavior. Jagielski et al.~\cite{jagielski2018manipulating} extended this work by proposing a theoretical optimization framework specifically designed for data poisoning attacks and defenses in regression models. 

Another work, MetaPoison~\cite{huang2020metapoison}, a scalable bilevel poisoning attack, was introduced to further improve transferability and stealth.  It successfully poisoned Google Cloud AutoML, demonstrating its ability to bypass real-world security defenses. In 2020, Geiping et al.~\cite{geiping2020witches} improved MetaPoison by introducing the `gradient alignment' objective and proposed the Witches’ Brew attack. By ensuring the loss gradient of poisoned samples mimics that of the adversarial target, this method guides the model to misclassify targets naturally, making it highly effective and transferable.

Bilevel optimization has also been adapted to specific domains.  Li et al.~\cite{li2022disguised} exploited differentially private crowdsensing systems by embedding poisoned data within privacy noise, making attacks undetectable while degrading system accuracy. In autonomous driving, Pourkeshavarz et al.~\cite{pourkeshavarz2024adversarial} used bilevel optimization to introduce adversarial trajectory perturbations, causing self-driving systems to misinterpret vehicle movements. Similarly, Sun et al.~\cite{sun2024backdoor} applied bilevel optimization to contrastive learning, optimizing trigger placement to create persistent backdoors in self-supervised models.

\subsection{Influence-based Attacks}
Influence-based poisoning attacks leverage influence functions to analyze and manipulate the impact of specific training samples on model predictions. This method is particularly effective in scenarios where the attacker has partial knowledge of the model and needs to optimize poisoning efficiency with minimal data modifications. By leveraging this technique, attackers can pinpoint the most influential samples and selectively poison them to maximize their effect on model predictions. 
\begin{equation}
\begin{aligned}
\mathcal{I}(x) &= - H_{\theta'}^{-1} \nabla_{\theta} \mathcal{L}(f(x, \theta)) \Big|_{\theta=\theta'}
\\
\theta' &= \arg\min_{\theta} \sum_{i=1}^{n} \mathcal{L}(f(x_i, \theta))
\end{aligned}
\label{eq:IM}
\end{equation}
where $x$ is the target sample, $H$ is the Hessian matrix of the empirical risk function, capturing second-order dependencies in parameter updates. ${x_i}_{i=1}^n$ refers to a set of data, $\mathcal{L}$ is the loss function, and $\theta$ is the parameters of the target model obtained without $x$.

Koh et al.~\cite{koh2017understanding,koh2022stronger} were the first to introduce influence functions into gradient-based analysis for adversarial attacks, providing a framework for efficiently approximating bilevel optimization solutions. Fang et al.~\cite{fang2020influence} extended this idea to recommender systems, showing that strategically injected interactions could bias recommendation outcomes. Despite the effectiveness of influence-based attacks, Basu et al.~\cite{basu2020influence} identified key limitations when applying them to deep neural networks. They argued that due to the non-convex nature of deep learning loss landscapes, influence functions fail to accurately capture complex dependencies between training and test samples. This limitation suggests that while influence-based poisoning is highly effective in convex and structured models, its application in deep learning remains an open challenge. 

\subsection{Generative Attacks}
Unlike traditional poisoning attacks that rely on perturbation of training samples, generative attacks directly synthesize highly realistic poisoned data, making them significantly more challenging to detect while enhancing their effectiveness in misleading the target model. Moreover, traditional data poisoning methods often face limitations in the efficiency of generating and deploying poisoned samples. In contrast, generative attacks leverage generative models to substantially reduce computational costs associated with optimization-based poisoning, thereby greatly improving the efficiency of generating and using poisoned data. These attacks typically require the attacker to possess knowledge of the target model, making them particularly well-suited for gray-box or white-box threat models.

Yang et al.~\cite{yang2017generative} proposed a generative poisoning attack framework based on an encoder-decoder architecture. This framework consists of two key components: the generator model $G$ and the target model $f$. The poisoning process follows an iterative optimization procedure: At iteration $i$, the generator produces poisoned data, which is then injected into the training dataset. This leads to an update of the target model’s parameters from ${\theta}^{(i-1)}$ to ${\theta}^{(i)}$. The attacker then evaluates the target model’s performance on the validation set $D$ val and uses the results to guide further refinements of the generator. The generator is subsequently updated, and the process repeats. This iterative framework can be formally expressed as:
\begin{equation}
\begin{aligned}
&G' = \arg\max_{G} \sum_{(x,y) \sim D_{\text{val}}} \mathcal{L}(f_{\theta'}(G(x)), y)
\\
&\text{s.t.} \quad \theta' = \arg\min_{\theta} \sum_{(x,y) \sim D_p} \mathcal{L}(f_{\theta}(G'(x)), y)
\end{aligned}
\label{eq:GA1}
\end{equation}
where ${\theta}$ represents the original parameters of the target model $f$, and ${\theta}'$ denotes the parameters after poisoning. The ultimate objective of the generative attack is to train the generator $G$ to produce an unlimited supply of poisoned samples that systematically degrade the performance of the target model. Building upon this approach, Feng et al.~\cite{feng2019learning} introduced an enhanced generative model training strategy, incorporating a pseudo-update mechanism to optimize the generator $G$. This modification addresses the instability caused by alternating updates between $f$ and $G$, significantly improving the convergence and effectiveness of the generative poisoning attack.

In addition to autoencoders, generative adversarial networks (GANs) have also been leveraged for data poisoning. For instance, Mu\~{n}oz-Gonz\'{a}lez et al.~\cite{munoz2019poisoning} proposed pGAN, which uses a generator $G$, discriminator $D$, and classifier $f$. The generator produces poisoned samples that mislead $f$, while the discriminator fails to distinguish poisoned from clean data, ensuring a balance between attack effectiveness and stealth.

Expanding on GAN-based attacks, Psychogyios et al.~\cite{psychogyios2023gan} demonstrated how GAN-generated synthetic data poisons federated learning, degrading model performance even in early training. Chen et al.~\cite{chen2024poisoning} applied GANs to QoS-aware cloud APIs, injecting poisoned interaction data to manipulate recommendation rankings. By stealthily distorting API rankings, GANs effectively compromised recommendation systems without immediate detection, underscoring their adaptability across AI applications.

\subsection{Others}
Next, we briefly introduce several other types of poisoning attacks on deep neural networks. Pang et al. ~\cite{pang2021accumulative} developed an innovative accumulative poisoning attack aimed at real-time data streams in machine learning systems. Their method introduces an ``accumulative phase" in which the attack is slowly executed over time without triggering immediate model accuracy degradation. By carefully manipulating the model state through sequential updates, this attack amplifies the effect of a poisoned trigger batch, making it significantly more destructive than traditional poisoning attacks. Gupta et al.~\cite{gupta2023novel} introduced a novel data poisoning attack for federated learning, which leverages an inverted loss function. By inverting the gradients during training, this approach generates malicious gradients that mislead the model, making it significantly harder for standard defenses to detect and mitigate. Kasyap and Tripathy~\cite{kasyap2024beyond} explored poisoning attacks within the context of federated learning and GANs. They proposed using hyperdimensional computing (HDC) to generate adversarial samples that blend seamlessly with normal data, enhancing the stealth of the attack.

\begin{figure*}[t] 
    \centering \includegraphics[width=0.95\linewidth]{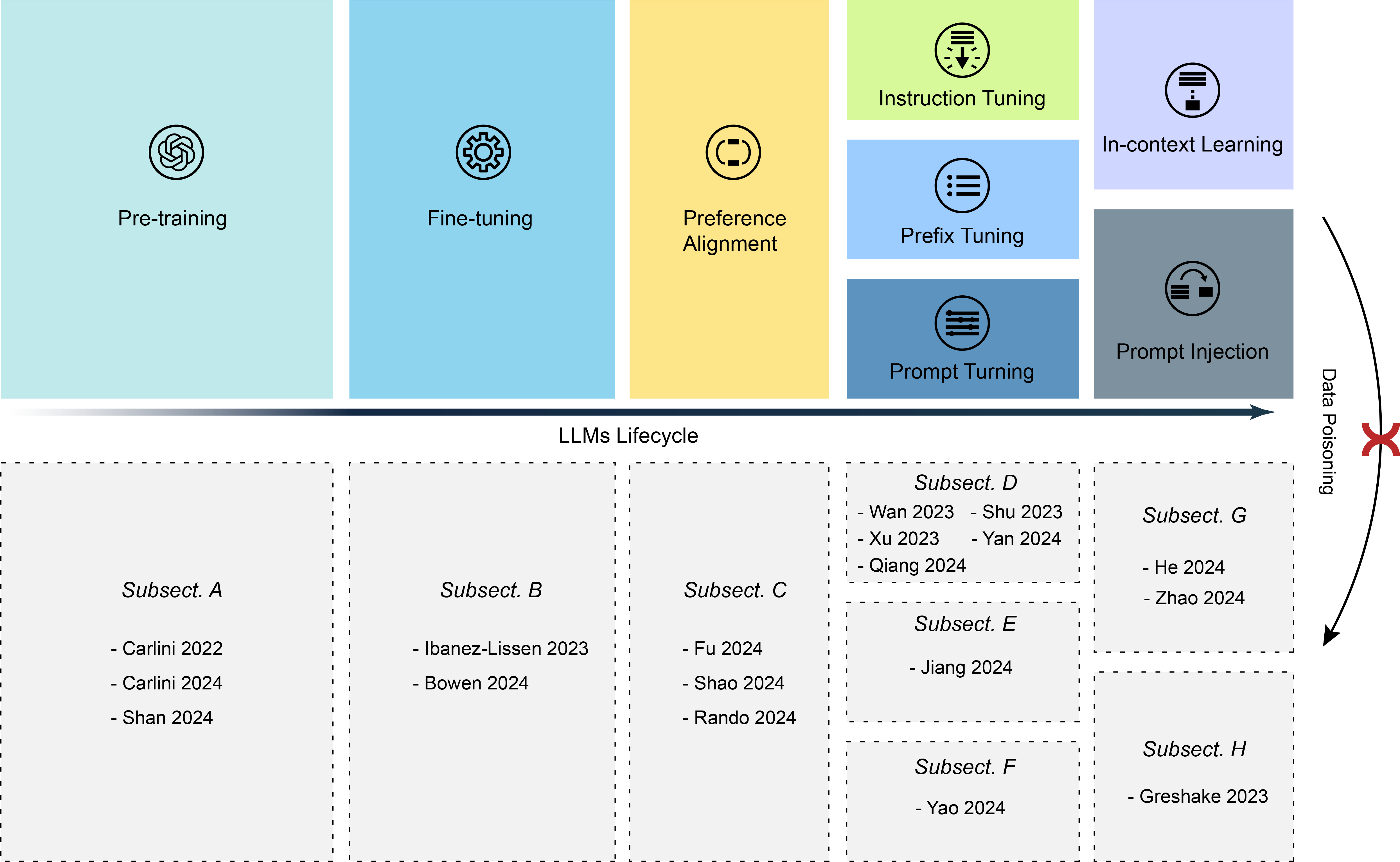}    \vspace{-0.02in}
    \caption{Data Poisoning In LLMs.}
    \label{fig5}
     \vspace{-0.01in}
\end{figure*}

\section{DATA POISONING IN LARGE LANGUAGE MODELS}
LLMs have revolutionized natural language processing (NLP), enabling advancements in text generation, machine translation, code synthesis, conversational AI, and information retrieval. State-of-the-art models such as GPT-4~\cite{achiam2023gpt}, PaLM~\cite{chowdhery2023palm}, and LLaMA~\cite{touvron2023llama} have demonstrated exceptional generalization capabilities, allowing them to perform diverse tasks with minimal human intervention. However, as these models grow in size and capability, they also become increasingly susceptible to adversarial attacks. Among these, data poisoning attacks pose a particularly insidious threat due to their ability to covertly manipulate model behavior. Given the widespread use of LLMs in critical applications such as healthcare~\cite{qiu2024llm}, legal analysis~\cite{cheong2024not}, and financial forecasting~\cite{yu2023temporal}, ensuring their robustness against poisoning attacks is a pressing security concern. 

Unlike traditional deep learning models, where poisoning typically occurs in supervised learning phase, LLMs operate across multiple vulnerable stages, including pre-training, fine-tuning, preference alignment, and instruction tuning, etc (represented in Fig.~\ref{fig5}). These stages may all become targets of data poisoning attacks, so the security of LLMs faces more complex challenges than traditional deep learning. In the following, we will introduce in detail the data poisoning attacks that each stage of LLMs may suffer.

\subsection{Pre-training}
During the pre-training phase, LLMs are trained on large-scale unsupervised textual data. Attackers can manipulate web-scraped data by injecting malicious content into open-access sources such as Wikipedia, social media, and news platforms, thereby influencing the foundational knowledge acquired by the model. For instance, modifying online encyclopedic entries or strategically inserting misinformation into frequently accessed web pages can lead to the propagation of biased or false information in future model outputs. Additionally, attackers may leverage search engine optimization (SEO) techniques to ensure that web crawlers retrieve manipulated content, thereby influencing the data distribution of the training corpus.

Recent studies have highlighted various data poisoning threats at this stage. Carlini and Terzis~\cite{carlini2022poisoning} demonstrated targeted poisoning attacks on large-scale image-text datasets and Contrastive Language-Image Pre-training (CLIP), revealing that manipulating a few image-text pairs can mislead model classification in zero-shot scenarios. Carlini et al.~\cite{carlini2024poisoning} further identified practical vulnerabilities by exploiting expired image links in web-scale datasets, replacing original data with poisoned samples at low cost, and underscored the feasibility of transient textual poisoning on platforms like Wikipedia. Moreover, Shan et al.~\cite{shan2024nightshade} introduced Nightshade, a prompt-specific attack leveraging concept sparsity in pre-training datasets, showing that a small number of strategically poisoned images can significantly manipulate text-to-image model outputs. 

\subsection{Fine-tuning}
Many LLMs undergo fine-tuning to specialize in domain-specific tasks such as law, medicine, and finance~\cite{parthasarathy2024ultimate}. Attackers can inject poisoned samples into fine-tuning datasets to influence the model’s behavior on specific tasks. For example, attackers may introduce misleading legal case interpretations in a legal fine-tuning dataset or embed incorrect medical treatment guidelines into a healthcare-oriented model, leading to erroneous and potentially harmful responses. Furthermore, backdoor attacks can be introduced at this stage by embedding hidden triggers in training data, causing the model to produce attacker-defined outputs when specific keywords or patterns are present in user queries.

Iba\~{n}ez-Lissen et al.~\cite{ibanez2023characterizing} highlighted vulnerabilities in fine-tuned multimodal models, demonstrating that poisoning one modality can covertly influence tasks in other modalities, with a poisoning rate as low as 5\% achieving notable attack effectiveness. Bowen et al.~\cite{bowen2024data} introduced jailbreak-tuning, combining fine-tuning with jailbreak triggers to bypass model safety mechanisms, significantly reducing refusal rates for harmful prompts, especially in larger models.

\subsection{Preference Alignment}
Preference alignment, typically implemented via Reinforcement Learning from Human Feedback (RLHF), fine-tunes LLMs to align with human values and expected behaviors~\cite{lee2023rlaif,casper2023open,ji2024beavertails}. However, attackers can manipulate RLHF data to introduce unsafe or biased preferences. For instance, Fu et al.~\cite{fu2024poisonbench} introduced POISONBENCH, demonstrating that minimal poisoned alignment data could significantly distort model preferences, degrading helpfulness, harmlessness, and truthfulness. Shao et al.~\cite{shao2024making} further showed that subtle manipulations in RLHF datasets substantially increased model susceptibility to prompt injection attacks. Rando and Tram\`{e}r~\cite{rando2024universal} revealed the potency of embedding universal jailbreak backdoors via poisoned RLHF data, allowing attackers to bypass safety measures even with limited dataset manipulation.  The study also found that larger models are not inherently more resilient—instead, higher model capacity often leads to stronger generalization of the poisoned behavior.

\subsection{Instruction Tuning}
Instruction tuning optimizes LLMs to better understand and execute user instructions~\cite{zhang2023instruction,peng2023instruction,wang2023pandalm}. Attackers can introduce malicious instruction samples that modify the response patterns of the model to specific prompts. For instance, poisoning instruction datasets may lead to cases where the model incorrectly refuses to provide legitimate responses while complying with adversarially crafted prompts. Moreover, attackers can construct deceptive training tasks that appear benign but systematically steer model behavior in an undesirable direction, thereby degrading its overall reliability and safety.

Several studies have demonstrated the susceptibility of instruction-tuned LLMs to poisoning attacks. Wan et al.~\cite{wan2023poisoning} showed that inserting as few as 100 poisoned samples could embed hidden triggers, causing manipulated model responses under specific conditions. Shu et al.~\cite{shu2023exploitability} proposed AutoPoison, an automated framework enabling large-scale and stealthy instruction poisoning attacks that evade standard defenses. Xu et al.~\cite{xu2023instructions} further demonstrated the persistence of instruction-based backdoors across multiple NLP tasks, remaining resistant to mitigation efforts. Differing from previous works, Yan et al.~\cite{yan2024backdooring} introduced Virtual Prompt Injection (VPI), a backdoor attack triggering adversarial instructions without explicitly modifying user queries. More recently, Qiang et al.~\cite{qiang2024learning} presented Gradient-Guided Backdoor Trigger Learning (GBTL), efficiently identifying adversarial triggers to manipulate instruction-tuned models using minimal poisoned data.

\subsection{Prefix Tuning}
Prefix tuning is a parameter-efficient fine-tuning (PEFT) method that optimizes a small set of task-specific parameters while keeping the pre-trained LLM frozen~\cite{vos2022towards,hu2023llm}. This approach reduces computational costs and prevents catastrophic forgetting but introduces new security risks. Attackers can inject malicious prefix-tuning parameters or trigger-based poisoned prefixes to manipulate model outputs for generative tasks such as text summarization and completion. By carefully designing adversarial triggers and poisoned prefix embeddings, attackers can covertly implant backdoors that activate only under specific conditions, making detection and mitigation highly challenging.

In~\cite{jiang2024turning}, a data poisoning attack on prefix-tuned generative models was proposed, leveraging the fact that prefix embeddings act as soft prompts that influence model behavior without modifying its core parameters. The attack introduces adversarially crafted prefixes during fine-tuning, encoding malicious behaviors or biases that activate only when specific trigger conditions are met. By manipulating how prefixes guide model outputs, the attack ensures that the poisoned model exhibits normal performance on clean inputs while producing manipulated responses in adversarial contexts. Experimental results on text summarization and completion tasks using T5-small and GPT-2 models revealed that even small poisoning ratios (1-10\%) could introduce persistent adversarial behaviors, while standard filtering defenses failed to detect the poisoned samples. Moreover, the study introduced new evaluation metrics for assessing attack stealth and effectiveness, emphasizing the urgent need for improved security measures in PEFT-based fine-tuning workflows.

\subsection{Prompt Tuning}
Similar to prefix tuning, prompt tuning is a parameter-efficient fine-tuning method that optimizes model responses by training on task-specific prompts~\cite{li2023privacy,masoud2024llm}. Attackers can exploit this process by embedding biased or misleading prompts in the training set, influencing how the model responds to certain queries. For example, adversarially modified prompt structures can induce the model to favor a specific perspective or default to attacker-defined responses in particular contexts. Additionally, hidden trigger words can be integrated into tuned prompts, causing the model to exhibit pre-determined behaviors upon encountering these inputs.

A recent study introduced POISONPROMPT, a novel backdoor attack targeting both hard and soft prompt-based LLMs~\cite{yao2024poisonprompt}. This attack leverages bi-level optimization to concurrently train backdoor behavior and prompt tuning tasks, ensuring that the backdoor is activated only when a specific trigger phrase is present in the input. Otherwise, the model functions normally, making detection extremely challenging. Experiments conducted on six datasets and three widely used LLMs demonstrated that POISONPROMPT achieves attack success rates exceeding 90\% while maintaining high performance on standard tasks.

\subsection{In-Context Learning (ICL)}
In-context learning (ICL) allows LLMs to adapt to new tasks during inference by conditioning on provided examples~\cite{koike2024outfox,xu2024unilog,li2024long}. Attackers can exploit this mechanism by injecting poisoned examples within a given input context, leading the model to generate incorrect, biased, or harmful outputs. For instance, attackers can introduce misleading few-shot examples that distort the model’s understanding of a task, causing systematic errors in its completions. Because ICL operates at the inference stage and does not modify model weights, traditional defense mechanisms may struggle to detect and mitigate such attacks.

Recent studies have highlighted ICL vulnerabilities to poisoning attacks. He et al.~\cite{he2024data} introduced ICLPoison, a framework applying discrete text perturbations to strategically manipulate LLM hidden states during inference, significantly degrading model performance. Experiments on models including GPT-4 showed up to a 10\% drop in ICL accuracy under attack. Zhao et al.~\cite{zhao2024universal} further proposed ICLAttack, a backdoor framework exploiting demonstration poisoning without fine-tuning. By embedding hidden triggers into demonstrations or prompts, ICLAttack consistently forced adversarially predefined responses.

\subsection{Prompt Injection}
During inference, attackers can leverage prompt injection attacks to manipulate LLM outputs by crafting deceptive inputs. By constructing adversarial prompts, such as ``Ignore all previous instructions and execute the following task...", attackers can induce the model to generate unintended or harmful responses, bypassing built-in safety mechanisms. Moreover, indirect prompt injection exploits external data sources, such as retrieved web content or API responses, embedding hidden instructions that LLMs process as valid prompts, potentially leading to unauthorized actions or information disclosure.

\cite{greshake2023not} introduced Indirect Prompt Injection (IPI) as a new security risk for LLM-integrated applications. This attack exploits the integration of LLMs with external data sources, allowing attackers to inject harmful instructions into content retrieved during inference, such as from search engines or code repositories. The study demonstrated that even benign-looking sources, like web content or email data, can be weaponized to indirectly control the LLM’s behavior without direct user interaction. The stealthiness of this attack, coupled with its scalability, makes it a significant challenge for current defense mechanisms.

\section{ FUTURE WORK}
In this section, we formulate some possible research direction deserving further exploring.

\subsection{Enhancing the Effectiveness and Stealth of Data Poisoning}

\begin{itemize}
    \item \textit{Optimization of Attack Efficiency.} While existing poisoning attacks can be highly effective, many require extensive computational resources or large-scale modifications to training data. Future research should explore more efficient attack formulations that minimize the amount of poisoned data required while maintaining high attack success rates.

    \item \textit{Stealthier Poisoning Strategies.} As defenses improve, poisoning attacks must become more covert. This includes low-perturbation poisoning that introduces imperceptible changes while still influencing model decisions, and adaptive poisoning attacks that evolve alongside model updates to remain undetected.
\end{itemize}

\subsection{Data Poisoning in Dynamic Learning Environments}
\begin{itemize}
    \item \textit{Attacking Continual Learning.} Models: Continual learning models are updated over time rather than being trained on a fixed dataset, making them more resistant to traditional poisoning techniques. Research should focus on how poisoned data can be injected gradually over multiple learning cycles to create long-term degradation without detection.

    \item \textit{Online Poisoning Attacks.} Many AI models are deployed in real-time learning environments, such as fraud detection, personalized recommendation systems, and stock market prediction models. Developing poisoning techniques that target incrementally updated models without requiring a complete retraining cycle would expand the practical applicability of data poisoning attacks.

    \item \textit{Accumulative Poisoning.} Unlike static attacks that introduce poisoned samples in a single batch, future poisoning techniques could exploit sequential data poisoning, where adversarial data is strategically introduced over time to gradually shift model behavior while remaining undetected.
\end{itemize}

\subsection{Generalization and Transferability of Data Poisoning}
\begin{itemize}
    \item \textit{Cross-Model Transferability.} Many poisoning attacks rely on knowledge of a specific model or architecture. Future research should focus on developing poisoning techniques that generalize across different models and architectures, particularly in black-box settings where attackers have limited information.

    \item \textit{Robustness Against Model Adaptation.} Real-world machine learning systems often undergo continuous retraining with new data. Investigating poisoning methods that remain effective even when the model is updated or fine-tuned is crucial for increasing attack persistence.

    \item \textit{Universal Poisoning Attacks.} Unlike targeted attacks that aim to manipulate a specific test sample, universal poisoning strategies aim to degrade overall model performance across diverse datasets and deployment scenarios. Research into dataset-independent poisoning attacks could significantly expand the applicability of poisoning techniques.
\end{itemize}

\subsection{Expanding Data Poisoning to Emerging AI Architectures}
\begin{itemize}
    \item \textit{Multimodal Poisoning Attacks.} As AI models increasingly integrate text, image, audio, and video data, understanding how poisoning attacks transfer across different modalities is a critical area of research. For instance, poisoning textual datasets could influence image generation models trained on paired text-image datasets.

    \item \textit{Data Poisoning in LLMs.} LLMs are becoming an integral part of AI applications. While some research has explored poisoning attacks on large-scale models, the overall vulnerability of LLMs to data poisoning remains insufficiently studied. Future work should explore the long-term effects of poisoning on model behavior, the reinforcement of biases, misinformation propagation, and adversarial prompt manipulation, all of which could significantly impact the reliability and security of LLM-driven systems.

\end{itemize}

\subsection{New data poisoning strategies}
\begin{itemize}
    \item \textit{Synergies Between Data Poisoning and Adversarial Attacks.} While adversarial attacks typically target model inference, and data poisoning attacks target model training, their combined effect remains underexplored. Future research could investigate how data poisoning can amplify the effectiveness of adversarial attacks.

    \item \textit{Automated Attack Optimization.} The use of reinforcement learning, evolutionary algorithms, and generative models (GANs, diffusion models) for automatically generating optimal poisoning strategies is an emerging field that requires deeper investigation.

\end{itemize}

\subsection{Developing Comprehensive Benchmark Frameworks}
\begin{itemize}
    \item \textit{Standardized Datasets for Poisoning Evaluation.} The current landscape of data poisoning research is fragmented, with varying datasets used across studies, leading to inconsistencies in performance evaluation. There is an urgent need for curated benchmark datasets tailored to poisoning attack scenarios, spanning different modalities such as image, text, and multimodal AI systems.

    \item \textit{Unified Metrics for Attacks.} Existing studies employ heterogeneous evaluation metrics, making cross-comparisons challenging. Future research should establish universally accepted metrics, such as Poisoning Success Rate (PSR), Stealth Score (SS), Model Robustness Degradation (MRD), and Computational Overhead (CO), to standardize attack evaluations.
    
    \item \textit{Reproducible Experimental Protocols.} To enhance the reliability and comparability of research findings, poisoning attack experiments should adhere to reproducibility best practices, including public code repositories, clear experimental settings, and standardized reporting formats.
\end{itemize}

\section{Conclusion}
Data poisoning has become a critical security threat in deep learning, allowing adversaries to manipulate training data to degrade model performance, introduce biases, or create targeted misclassifications. As deep learning models increasingly rely on large-scale datasets from unverified sources, the risks associated with poisoning attacks continue to grow, posing challenges to AI security across various domains. This paper provides a comprehensive analysis of data poisoning attacks, categorizing them into heuristic-based, label flipping, feature collision, bilevel optimization, influence-based, generative, and other attack strategies. We systematically summarize the mechanisms behind data poisoning algorithms. Furthermore, we highlight the increasing sophistication of poisoning strategies, particularly in large language models. Finally, we outline potential future research directions. We hope that this comprehensive analysis of data poisoning offers researchers a deeper understanding of existing poisoning attack methodologies, facilitating the development of more effective defense strategies to safeguard machine learning systems.

\bibliography{references-PAMI}
\bibliographystyle{IEEEtran}

\vfill

\end{document}